\documentclass[preprint,aps,longbibliography]{revtex4-1}
\usepackage{amsfonts,amssymb,amsmath,bm}
\usepackage[dvips]{graphicx}
\usepackage{color}
\usepackage[dvipsnames]{xcolor}
\usepackage{hyperref}
\usepackage{times}
\usepackage{rotating}

\definecolor{bostonuniversityred}{rgb}{0.8, 0.0, 0.0}
\definecolor{dukeblue}{rgb}{0.0, 0.0, 0.61}

\renewcommand{\Re}{\mathrm{Re}}

\begin{document}

\title{Formation and decay of eddy currents generated by crossed surface waves}

\author{V. M. Parfenyev$^{1,2}$}
\author{S. V. Filatov$^{3}$}
\author{M. Yu. Brazhnikov$^{2,3}$}
\author{S. S. Vergeles$^{1,2}$}
\author{A. A. Levchenko$^{3}$}

\affiliation{$^1$Landau Institute for Theoretical Physics, Russian Academy of Sciences, 1-A Akademika Semenova av., 142432 Chernogolovka, Russia}
\affiliation{$^2$National Research University Higher School of Economics, Faculty of Physics, Myasnitskaya 20, 101000 Moscow, Russia}
\affiliation{$^3$Institute of Solid State Physics, Russian Academy of Sciences, 2 Academician Ossipyan str., 142432 Chernogolovka, Russia}

\date{\today}

\begin{abstract}
The mass-transport induced by crossed surface waves consists of the Stokes and Euler contributions which are very different in nature. The first contribution is a generalization of Stokes drift for a plane wave in ideal fluid and the second contribution arises due to the fluid viscosity and it is excited by a force applied in the viscous sublayer near the fluid surface. We study the formation and decay of the induced mass-transport theoretically and experimentally and demonstrate that both contributions have different time scales for typical experimental conditions.
The evolution of the Euler contribution is described by a diffusion equation, where the fluid kinematic viscosity plays the role of the diffusion coefficient, while the Stokes contribution instantly tracks the wave pattern and therefore it evolves faster due to the additional wave damping near the system boundaries.
The difference becomes more pronounced if the fluid surface is contaminated. We model the effect of contamination by a thin insoluble liquid film presented on the fluid surface with the compression modulus being the only non-zero rheological parameter of the film. Then the Euler contribution into the mass-transport becomes larger and the evolution of the Stokes contribution becomes faster as compared to the free surface case.
We infer the value of the compression modulus of the film by fitting the results of transient measurements of eddy currents and demonstrate that the obtained value leads to the correct ratio of amplitudes of horizontal and vertical velocities of the wave motion and is in reasonable agreement with the measured dissipation rate of surface waves.
\end{abstract}

\maketitle

\section{Introduction}

The horizontal transport of Lagrangian particles in a fluid produced by surface waves is a long-standing problem of both fundamental and practical interest. The first attempt to explain this phenomenon dates back to the classical paper by George Stokes \cite{stokes1847theory}, in which he investigated the problem for the irrotational progressive wave in an ideal fluid. He showed that the Lagrangian particles possess a second-order (with respect to the wave amplitude) drift velocity, which is now called the Stokes drift. Later, Michael Longuet-Higgins found that the fluid viscosity breaks the irrotational approximation and substantially modifies the drift velocity \cite{longuet1953mass}.

The influence of the fluid viscosity on the transport of Lagrangian particles can be explained as follows. A surface wave possesses a momentum that is directed parallel to the direction of propagation and is proportional to the square of the wave amplitude. Viscous dissipation leads to a decrease in the amplitude of the wave during its propagation. It means that the momentum associated with the wave motion also decreases. Then the conservation of the total momentum requires the presence of a force acting on the fluid. This force is applied near the fluid surface (in the viscous sublayer) and it is of the second order in the wave amplitude and linear in the viscosity \cite{weber2001virtual}. The action of this force leads to the generation of a slow (second-order) current, which then spreads into the fluid bulk due to the viscosity. In the stationary regime, the drift velocity associated with the slow current is independent of the fluid viscosity, even though it originates from the viscosity. The phenomenon is very similar to the acoustic streaming produced in a fluid during the propagation of a sound wave \cite{boluriaan2003acoustic}.

Recently, interest in this problem has appeared again, but in the more complex formulation. How will the mass-transport be arranged if crossed waves are excited on the fluid surface \cite{filatov2016nonlinear, francois2017wave}? Theoretical analysis of the stationary regime shows that in the case of excitation of monochromatic standing perpendicular waves on the surface of deep fluid, a regular lattice of eddy currents is formed and its period is determined by the wavelength \cite{parfenyev2018influence}. The eddy currents can be described as a sum of Stokes drift and Euler contribution, which takes into account the current originating from the fluid viscosity. Both terms have the same horizontal structure and decay exponentially with depth, but the decrements are numerically different: the Stokes drift decreases faster. Let us note that if the fluid surface is free then the Euler contribution is independent of the fluid viscosity. This conclusion agrees well with the results reported in Ref.~\cite{longuet1953mass} and briefly discussed above.

The problem becomes more complex if one tries to take into account the possible presence of a surface film, for example, due to various contaminants and impurities \cite{kang1995steady}. In particular, in Ref.~\cite{parfenyev2018influence} we show that a thin insoluble liquid (with zero shear elasticity) film substantially changes the Euler contribution to the drift velocity of Lagrangian particles as compared to the free surface case. Now it depends on the fluid viscosity and compressibility properties of the surface film. Let us stress that these changes occur not only in a thin viscous sublayer near the surface but also in the fluid bulk.

In this paper, we extend the theoretical description to the non-stationary processes of decay and formation of eddy currents and present experimental results which are in quantitative agreement with the proposed model. We study formation and decay of eddy currents generated by crossed waves on the fluid surface, which could be contaminated for typical experimental conditions, see, e.g., Ref.~\cite{campagne2018impact}. We model the effect of contamination by a presence of a thin liquid film on the fluid surface, and based on transient measurements of the wave elevation, we theoretically obtain the evolution of the intensity of eddy currents and then compare it to the experimental data. By fitting the experimental data, we infer the elastic modulus of the surface film which is the only parameter that characterizes its properties in our model. We demonstrate that the obtained value of the elastic modulus leads to the correct ratio of amplitudes of horizontal and vertical velocities of the wave motion and is in reasonable agreement with the measured dissipation rate of surface waves, both of which are modified due to surface contamination. The obtained results allow one to separate the Stokes drift and the Euler contribution, confirm the correctness of the description of eddy currents generated by crossed surface waves presented in Ref.~\cite{parfenyev2018influence} and extend the theoretical description to the non-stationary processes.

\section{Theoretical Model}

%In that way one can obtain the explicit expression for the slow current ${\bm V}_E$ in terms of the excited wave motion, see, e.g., Refs.~\cite{longuet1953mass,filatov2016nonlinear,parfenyev2018influence}.

The wave motion (relatively fast) excited on the surface of viscous fluid necessarily causes the generation of a current (relatively slow) due to hydrodynamic nonlinearity. Therefore, the velocity field of fluid in the Eulerian description can be represented as ${\bm v} = {\bm u} + {\bm V}_E$, where ${\bm u}$ and ${\bm V}_E$ correspond to the wave motion and the slow current respectively. In what follows, we consider gravity or capillary surface waves excited on the surface of fluid that is possibly covered by a thin insoluble liquid film. We assume that the slow current is weak, so the effective Reynolds number characterizing the current is small and one can analyze the hydrodynamic nonlinearity using perturbation theory. Then, describing the current, we can neglect the hydrodynamic self-nonlinearity and take into account only the pair nonlinear interaction between the excited surface waves, which leads to the generation of the current. In that way one can obtain the explicit expression for the slow current ${\bm V}_E$ in terms of the excited wave motion \cite{longuet1953mass,parfenyev2018influence}. A natural way to study the slow current experimentally is examining the trajectories of passive particles advected by the flow (induced mass-transport). The corresponding Lagrangian velocity ${\bm V}$ averaged over fast wave oscillations contains an additional contribution ${\bm V}_S$ associated with the Stokes drift mechanism, i.e. ${\bm V} = {\bm V}_E + {\bm V}_S$.
%The explicit stationary expressions for both contributions in terms of the excited wave motion was found in Refs.~\cite{longuet1953mass,parfenyev2018influence}.

Further, motivated by recent experimental studies \cite{filatov2016nonlinear, francois2017wave}, we consider the specific case when two orthogonal monochromatic standing waves are excited on the surface of fluid. The surface elevation $h(t,x,y)$ is given by
\begin{equation}\label{eq:1}
  h(t,x,y) = H_1 \cos(\omega t) \cos (kx) + H_2 \cos(\omega t + \psi) \cos (ky),
\end{equation}
where $H_1$ and $H_2$ are the amplitudes of the waves, $k$ is the wave number, $\omega$ is the wave frequency and $\psi$ is the phase shift between excited waves. The deep-water assumption is implied and we assume that the wave steepness is small, $kH_{1,2} \ll 1$, and that the waves are weakly decaying if the wave excitation is turned off, i.e. $\gamma = \sqrt{\nu k^2/\omega} \ll 1$, where $\nu$ is the fluid kinematic viscosity coefficient. The nonlinear interaction between these waves leads to the generation of slow currents, which form a regular lattice of horizontal vortices with a period determined by the wavelength. It is convenient to describe the corresponding mass-transport in terms of the vertical vorticity, $\Omega = \partial_x V_y - \partial_y V_x$, where $V_x$ and $V_y$ are horizontal components of the Lagrangian velocity of fluid particles. In Ref.~\cite{parfenyev2018influence} for the stationary regime we found the following result:
\begin{eqnarray}\label{eq:W}
&\Omega= \displaystyle \left[\frac{\varepsilon^2 e^{kz\sqrt{2}}}{2 \gamma (\varepsilon^2-\varepsilon\sqrt{2}+1)} + \sqrt{2} e^{kz\sqrt{2}} + e^{2kz} \right] \Lambda (x,y), \quad z \leq 0,&\\[5pt]
\label{eq:Lambda}
&\Lambda (x,y) = - H_1 H_2 \omega k^2 \sin (kx) \sin (ky) \sin \psi.&
\end{eqnarray}
Here $\varepsilon \geq 0$ is the dimensionless compression modulus of a thin film, which possibly covers the fluid surface. The axis $z$ is directed vertically, opposite to the gravitational acceleration, and $z=0$ corresponds to the unperturbed (without waves) fluid surface. The limiting case of a free surface corresponds to $\varepsilon \rightarrow 0$ and in the opposite case $\varepsilon \rightarrow \infty$ we deal with an almost incompressible surface film.
%\textcolor{blue}{Note that the implied deep-water assumption is fulfilled when the fluid depth is greater than the inverse wave number $1/k$ of the excited waves.}

A thin film on the fluid surface was introduced to model the effect of surface contamination, which takes place for typical experimental conditions, see, e.g., Ref.~\cite{campagne2018impact}. In general, the rheological properties of the film can be characterized by four coefficients: dilational elasticity, dilational viscosity, shear elasticity, and shear viscosity \cite{langevin2014rheology}. In our model, we assumed that the dissipation due to internal viscosity of the film is small as compared to the dissipation in the fluid bulk and therefore we neglected the dilational and shear viscosities of the film. The approximation is justified when $\eta \gg \eta_s k$, and here $\eta_s$ stands for the dilational/shear viscosity of the film and $\eta$ is the dynamic viscosity of the fluid. We also assumed that the film is liquid, i.e. it does not resist the shear deformations in the film plane and thus the shear elasticity is zero. Finally, we adopted that the film is formed by an insoluble agent and for this reason the film mass is conserved. In this way, we describe the film rheological properties by the only parameter --- the dilational elasticity or the compression modulus. As one can see, our consideration is limited to a rather narrow class of surface films and in this sense, our model is not universal. At the same time, it is simple and, as we will see, explains the experimental data quite well.

Since a film is formed on the fluid surface due to contamination, its compression modulus $\varepsilon$ is \textit{a priori} unknown. There are two methods to infer the value of $\varepsilon$ by analyzing the stationary motion of a surface wave, see Ref.~\cite{parfenyev2018influence}. In the first method, one needs to measure the amplitudes of horizontal $||u_{\alpha}||$ and vertical $||u_z||$ velocities on the fluid surface, and then calculate their ratio:
\begin{equation}\label{eq:4}
  \frac{||u_{\alpha}||}{||u_z||}
  =\displaystyle \frac{1}{\sqrt{\varepsilon^2-\varepsilon \sqrt{2}+1}}.
\end{equation}
Hereinafter $\alpha = \{x, y\}$ for the wave propagating in the $X-$ and $Y-$direction correspondingly. Note that for a free surface ($\varepsilon \rightarrow 0$) the maximum values of horizontal and vertical velocities on the fluid surface are equal to each other (well-known result for deep-water waves), while in the case of an almost incompressible film ($\varepsilon \rightarrow \infty$) the horizontal velocity on the fluid surface is zero.

The second method relates the compression modulus $\varepsilon$ of the surface film with the wave attenuation rate $1/\tau$ after the pumping is turned off:
\begin{equation}\label{eq:W''}
 \frac{1}{\omega \tau}
 =  2\gamma^2 \left(1 + \dfrac{1}{\gamma kL \sqrt{2}} \right) +\frac{\gamma}{2\sqrt{2}}\frac{\varepsilon^2}{(\varepsilon^2-\varepsilon\sqrt{2}+1)}.
\end{equation}
Here $L$ is the length of the side of the square cell which is used in the experiment and the corresponding term takes into account the dissipation near the system boundaries \cite[\S25]{landau1987course}. Other terms describe dissipation in the case of a borderless system \cite[Eq.(25)]{parfenyev2018influence}. Let us note that the presence of a thin film on the fluid surface changes only the wave damping; the dispersion law of surface waves remains the same, $\omega^2 = gk + \sigma k^3/\rho$, except for the possible change in an equilibrium value $\sigma$ of the surface tension coefficient. Hereinafter we denote the absolute value of gravitational acceleration by $g = 9.8 \, m/s^2$.

Above we have discussed that the mass-transport generated due to the nonlinear interaction of surface waves can be described as a sum of Stokes drift and Euler contribution, see also Ref.~\cite{parfenyev2018influence}. The Stokes drift and the Euler contribution are very different in nature. The Stokes drift is the result of nonlinear Lagrangian dynamics during one time period of oscillations and it does not produce any contribution into the mean velocity of fluid in the Euler description. In the stationary regime, the Stokes drift corresponds to the last term in expression (\ref{eq:W}), which is proportional to $\exp(2kz)$. If the amplitudes $H_{1,2}(t)$ and the phase difference $\psi(t)$ in expression (\ref{eq:1}) are time-dependent now, being slightly changed during one oscillation period, then the Stokes contribution instantly tracks these changes and everything one needs to do is to substitute $\Lambda(x,y)$ by
\begin{equation}\label{eq:6}
\Lambda(x,y,t) = - H_1 (t) H_2 (t) \omega k^2 \sin (kx) \sin (ky) \sin \psi(t).
\end{equation}
In contrast, the Euler contribution corresponds to the mean velocity of fluid. In the stationary regime, it is given by two first terms in expression (\ref{eq:W}), which both are proportional to $\exp(kz\sqrt{2})$. Their non-stationary behavior is non-trivial and it is the focus of this article. The Euler contribution is excited by a force, which is localized in the narrow viscous sublayer near the fluid surface and is produced due to hydrodynamic nonlinearity. Therefore, the dynamics of this contribution is relatively slow and it is determined by the fluid viscosity and inertia.

The exact equation which describes the dynamics of the Euler contribution was obtained in Ref.~\cite[Sec. IV]{parfenyev2018influence}. Since the exciting force is localized in the viscous sublayer of thickness $\delta \sim \gamma/k$ and $\delta \ll 1/k$, one can assume that the force is a tangent stress applied to the fluid surface at $z=0$ (it is also known as the virtual wave stress, see Ref.~\cite{longuet1969nonlinear}). This simplification does not change the solution of the equation in the fluid bulk at a depth $|z| \gg \delta$. Therefore, denoting the Euler contribution as $\Omega_E(x,y,z,t)$, we obtain the following equation
\begin{equation}
\label{eq:X0}
\partial_t \Omega_E - \nu \nabla^2 \Omega_E = 0, \qquad z<0,
\end{equation}
which has to be supplemented by a fixed-stress boundary condition at the surface and by the condition of the absence of eddy currents at infinite depth,
\begin{equation}\label{eq:X1}
  \nu \partial_z \Omega_E\big\vert_{z=0} = F \Lambda (x,y,t), \qquad \Omega_E\big\vert_{z \rightarrow -\infty} \rightarrow 0,
\end{equation}
see also Ref.~\cite[Eq.~(2.11)]{longuet1969nonlinear} and Ref.~\cite[Eq.~(14)]{parfenyev2018influence}. The diffusion equation with a fixed flux through the boundary and with a fixed source at the boundary are equivalent to each other, see Appendix~\ref{sec:AppA} for details. Thus, instead of the boundary-value problem (\ref{eq:X0})-(\ref{eq:X1}), one can solve the equation
\begin{equation}
\label{eq:X}
\partial_t \Omega_E - \nu \nabla^2 \Omega_E = 2 F \delta(z) \times \Lambda (x,y,t), \qquad -\infty<z<+\infty,
\end{equation}
with the boundary condition $\Omega_E \to 0$ as $z \to \pm \infty$ and with an initial condition that is symmetrically reflected from the plane $z=0$. Here $\delta(z)$ is the Dirac delta function which means that the excitation force is applied to the fluid surface, and $F$ characterizes the intensity of this force and its value is discussed below.

Strictly speaking, the solution of the presented boundary-value problem (\ref{eq:X}) may differ from the solution of the exact problem in the viscous sublayer near the fluid surface. However, in Ref.~\cite{parfenyev2018influence} it was shown that the corresponding contribution is canceled by the contribution to the Stokes drift, which is produced by the vortical corrections to the velocity field owing to the fluid viscosity and the presence of a surface film. As a result, the Stokes drift must be calculated taking into account only the potential contribution to the velocity field and the presented boundary-value problem (\ref{eq:X}) gives the correct result for the Euler contribution $\Omega_E(t)$ everywhere.

In principle, the source strength $F$ can be found by integrating the exact equations over a viscous sublayer. However, we already know that the time asymptotic value of $\Omega_E(t)$ in the case of stationary wave motion is equal to the sum of two first terms in expression (\ref{eq:W}). This means that the source strength should be equal to
\begin{equation}\label{eq:Z}
  F = 2 \nu k \left[1 + \frac{\varepsilon^2}{2 \sqrt{2} \gamma (\varepsilon^2-\varepsilon\sqrt{2}+1)} \right],
\end{equation}
see also equation (\ref{eq:formation}) below and the note after it. The system of equations (\ref{eq:X})-(\ref{eq:Z}) completely describes the evolution of Euler's contribution to the mass-transport provided that the wave motion is known. Note that the presence of a surface film increases the source strength $F$.

Now we would like to consider the following scenario which will be studied experimentally below. Suppose that initially the fluid was at rest, and then the monochromatic pumping which excites the surface waves is turned on at time $t=0$. The excited waves induce the slow current which consists of Euler and Stokes contributions and we wait for the establishment of a steady state. Then the pumping is turned off at time $t=t^\ast$ and we observe the decay of surface waves and the slow current due to the fluid viscosity.

Based on equation (\ref{eq:X}), one can conclude that the characteristic time $t_E$ of the evolution of the Euler contribution is the viscous diffusion time, $t_E = 1/(2 \nu k^2)$, and it does not feel the presence of surface film and system boundaries. On the contrary, the presence of the surface film and the friction against system boundaries increase the decay rate $1/\tau$ of the waves, see expression (\ref{eq:W''}). Thus, the characteristic time of the evolution of the Stokes drift correction to the mass-transport is smaller being estimated as $t_S = \tau/2$, since the effect is quadratic with respect to wave amplitudes. In the case of unbounded fluid with free surface, both times $t_E$ and $t_S$ are of the same order. Then it is necessary to measure the time dependence of the amplitudes of the waves $H_{1,2}(t)$ and the phase difference $\psi(t)$ and solve equation (\ref{eq:X}) numerically. However, for some experimental conditions, an additional wave dissipation due to the presence of the film and the walls can lead to the separation of the characteristic time scales, i.e. $t_E \gg t_S$. Then the right-hand side of equation (\ref{eq:X}) can be considered as nonzero constant for $0<t<t^\ast$ and equal to zero outside this time interval. In this case the equation can be solved analytically and below we present a solution.

First, we consider the process of formation of the Euler contribution. The initial condition is trivial, $\Omega_E(x,y,z,0) = 0$, and the right-hand side of equation (\ref{eq:X}) is fixed in time and space. To solve the equation we perform the Fourier transformation, $\displaystyle \tilde{\Omega}_E (x,y,q,t) = \int_{-\infty}^{\infty} dz \; \Omega_E (x,y,z,t) e^{-iqz}$,
and then we find
\begin{equation}\label{eq:11}
  \partial_t \tilde{\Omega}_E - \nu ((\partial_x^2+\partial_y^2)-q^2) \tilde{\Omega}_E = 2F \Lambda(x,y)
\end{equation}
with the initial condition $\tilde{\Omega}_E (x,y,q,0) = 0$. Since the horizontal spatial structure of $\Omega_E$ coincides with that of $\Lambda(x,y)$, see expression (\ref{eq:Lambda}), one can replace $\partial_x^2+\partial_y^2$ with $-2k^2$ in equation (\ref{eq:11}). Then the solution of this equation is
\begin{equation}\label{eq:formal-solution}
  \tilde{\Omega}_E (x,y,q,t)= 2F \Lambda(x,y) \int_{0}^{t} dt' \; e^{-\nu(2k^2+q^2)(t-t')},
\end{equation}
and after the inverse Fourier transformation, $\displaystyle \Omega_E (x,y,z,t) = \int_{-\infty}^{\infty} \dfrac{dq}{2\pi} \; \tilde{\Omega}_E (x,y,q,t) e^{iqz}$, we obtain:
\begin{equation}\label{eq:formation}
  \Omega_E (x,y,z,t) = \dfrac{F \Lambda(x,y)}{\sqrt{2\pi} \nu k} \int\limits_{0}^{2\nu k^2 t} \dfrac{d\xi}{\sqrt{\xi}} \; \exp \left(- \xi - \frac{(kz)^2}{2 \xi} \right).
\end{equation}
Note that the time asymptotic value, $t \rightarrow \infty$, corresponds to the sum of first two terms in equation (\ref{eq:W}), only if the source strength $F$ is given by expression (\ref{eq:Z}). The value of the Euler contribution on the fluid surface is of particular interest because it is relatively easy to measure experimentally. By substituting $z=0$ in expression (\ref{eq:formation}) and using relation (\ref{eq:Z}), we find:
\begin{equation}\label{eq:13}
\Omega_E (x,y,0,t)= \displaystyle \left[\frac{\varepsilon^2 }{2 \gamma (\varepsilon^2-\varepsilon\sqrt{2}+1)} + \sqrt{2} \right] \Lambda (x,y) \times \mathrm{Erf}(\sqrt{2\nu k^2 t}),
\end{equation}
where $\mathrm{Erf}(x) = \displaystyle \dfrac{2}{\sqrt{\pi}} \int_{0}^{x} d\zeta \; e^{-\zeta^2}$. Therefore, at the initial stage of formation, $t \ll t_E$, one obtains $\Omega_E (x,y,0,t) \propto \sqrt{t/t_E}$.

Next, we consider the process of decay of the Euler contribution after the pumping is turned off and the wave motion disappears, $t>t^\ast$. The initial condition is
\begin{equation}
  \Omega_E(x,y,z,t^\ast) = \left[\frac{\varepsilon^2 }{2 \gamma (\varepsilon^2-\varepsilon\sqrt{2}+1)} + \sqrt{2} \right] e^{-k|z|\sqrt{2}} \Lambda (x,y)
\end{equation}
and the evolution is governed by equation (\ref{eq:X}) with the right-hand side equals to zero. The equation can be easily solved in the Fourier space, and after the inverse transformation we obtain:
\begin{equation}\label{eq:decay}
  \Omega_E(x,y,z,t) = \dfrac{F \Lambda(x,y)}{\pi \nu} e^{-2\nu k^2 (t-t^\ast)} \int_{-\infty}^{+\infty} dq \; \dfrac{\cos(qz)}{2k^2+q^2} e^{-\nu q^2 (t-t^\ast)}.
\end{equation}
By substituting $z=0$ in expression (\ref{eq:decay}) and using relation (\ref{eq:Z}), we find the evolution of the Euler contribution on the fluid surface:
\begin{equation}\label{eq:16}
\Omega_E (x,y,0,t)= \displaystyle \left[\frac{\varepsilon^2 }{2 \gamma (\varepsilon^2-\varepsilon\sqrt{2}+1)} + \sqrt{2} \right] \Lambda (x,y) \times \left( 1 - \mathrm{Erf}(\sqrt{2\nu k^2 (t-t^\ast)}) \right).
\end{equation}
Thus, at the initial stage $(t-t^\ast) \ll t_E$, the decay is described by a square root law $\Omega_E (x,y,0,t) \propto \left(1- \sqrt{8 \nu k^2 (t-t^\ast)/\pi} \right)$, and at large times $(t-t^\ast)\gg t_E$ it turns to the exponential law $\Omega_E (x,y,0,t) \propto \sqrt{t_{E}/(t-t^\ast)}\exp(-(t-t^\ast)/t_{E})$ .

\section{Experiment}

\begin{figure}[t]
  \includegraphics[width=0.8\linewidth]{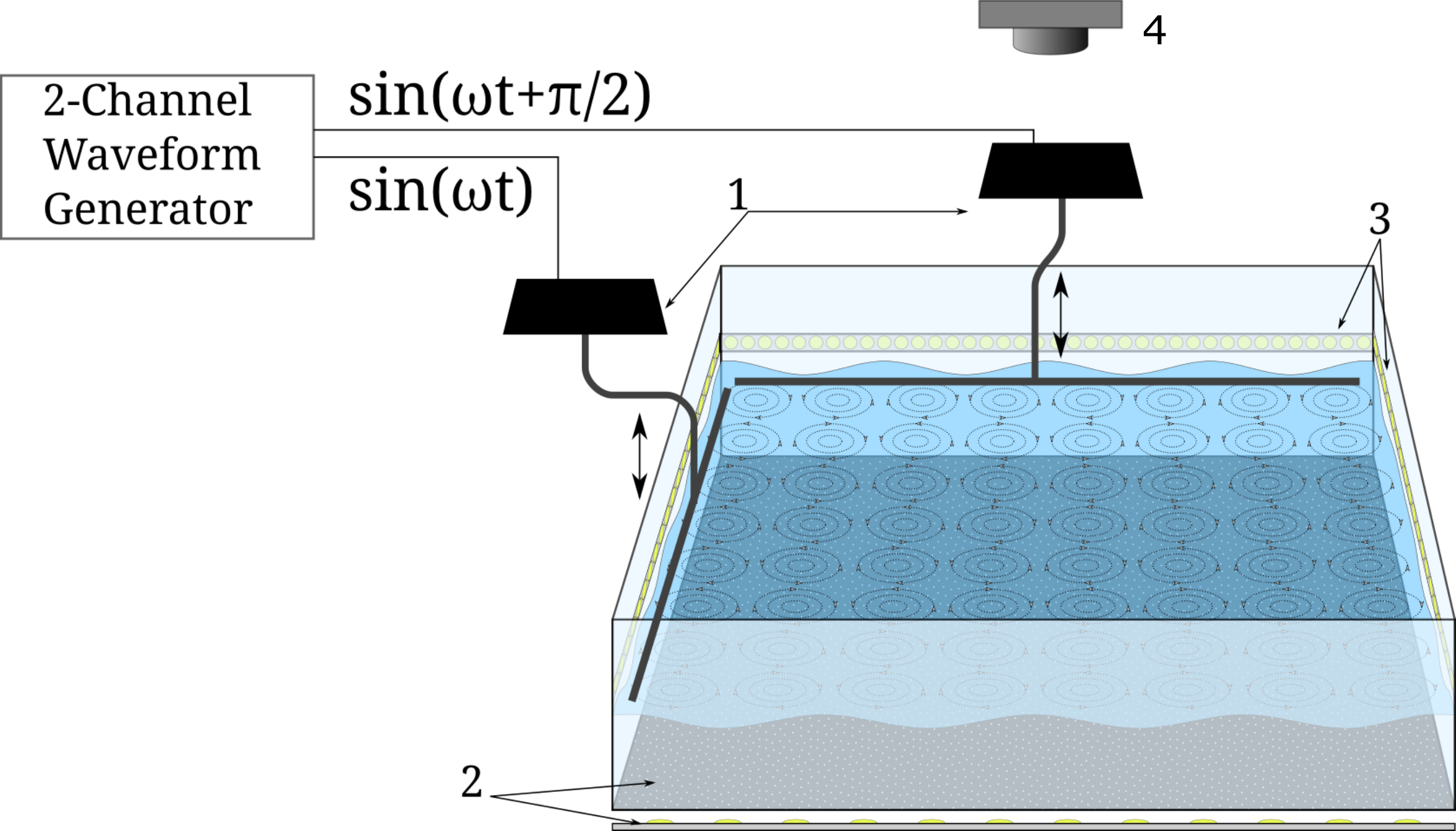}
  \caption{The scheme of the experimental setup: 1~-- plungers and driving mechanisms, 2~-- contrast pattern and bottom LED array used to highlight the pattern,  3~-- LED strips along the bath perimeter for illumination of tracers on the fluid surface, 4~-- photo camera that captures slow eddy currents and waves on the fluid surface.}
\label{pic:0}
\end{figure}

The formation and decay of eddy currents were studied experimentally in a square glass bath with the side length $L=70\, cm$ and depth of $20 \, cm$, see Fig.~\ref{pic:0}. The bath was installed on a Standa table with a pneumatic vibration-isolating suspension system. Surface waves were excited by two wave generators consisting of a plungers and driving mechanisms, which were installed on support frames mounted near two adjacent bath sides. The plungers were made of a stainless steel tubes closed on the both ends. The diameter of the plungers was $10 \, mm$, the length was $68 \, cm$. In the equilibrium the plunges were submerged into the fluid down to half of their diameter. Two TS-W254R subwoofers (Pioneer) with a nominal power of $250 \, W$ each were used to drive the plungers. Sinusoidal signals were generated by an Agilent $33500B$ two-channel generator, amplified and supplied to the subwoofers.

The theoretical description presented above and in Ref.~\cite{parfenyev2018influence} considers the hydrodynamic nonlinearity in a perturbative manner. It means that the advection by the slow currents should not destroy the vortex lattice, i.e. the advection time by the slow currents $t_{\mathrm{adv}}$ at scale of the order of the wavelength should be larger than the viscous diffusion time $t_E$ already defined at the same scale. If the only contribution to the slow currents is the vortex lattice then the requirement is equivalent to a small value of the effective Reynolds number, $\mathrm{Re} = \Omega/(\nu k^2) \lesssim 1$. In order to reduce the Reynolds number, the bath was filled with glycerine-water solution. The fraction of glycerine by mass was varied in the range from $0 \%$ to $64 \%$. The surface waves were excited at frequency of $3 \, Hz$. This excitation frequency corresponds to the surface gravity wave with the wave number $k=0.36 \pm 0.01 \, cm^{-1}$ (evaluated from the dispersion law $\omega^2 = gk + \sigma k^3/\rho$), which is very close to eigenmode inasmuch as $kL/\pi=8.02\pm0.02$. The uncertainty in the wave number accounts for the difference of the values of the surface tension coefficients and the mass densities of water and glycerine-water solutions used in the experiments. The fluid depth was $10 \, cm$ and therefore the adopted deep water assumption is fulfilled.

The velocity of eddy currents on the fluid surface was derived using particle image velocimetry method \cite{adrian1991particle}. A polyamide white powder with an average granule diameter of about $30 \, \mu m$ was poured on the fluid surface. The floating tracer particles on the surface were illuminated by LEDs fixed along the bath perimeter. Tracers motion was recorded by an EOS 70D camera located approximately $1.5 \, m$ above the fluid surface with a frequency of $24$ frames per second, which is multiple of the excitation frequency. Such a frequency made it possible to eliminate the oscillating component of the motion of a tracer particle on the surface at the excitation frequency by choosing every eighth pictures of the vibrating surface. The cross-correlation analysis of the image pairs using the PIVLab software \cite{thielicke2014pivlab} allowed us to obtain the velocity field associated with the tracer motion and then calculate the vertical vorticity $\Omega$ on the surface.

The fluid surface oscillations in the vertical direction were detected using recently developed technique~\cite{filatov2018technique}, which is based on reconstruction of the surface curvature by analyzing the optical distortion of a contrast image at the bottom of the bath. Pattern of clear dots randomly positioned against a dark lightproof background was printed on transparent film. LED placed under the transparent bottom of the bath illuminated the pattern and produced a picture of bright speckles with density of $2 \times 10^4 $ per square metre that was captured from the top by the same photo camera. Particle image velocimetry was used to determine displacement of the speckles on the pair of consecutive frames, which is proportional to the local slope of the fluid surface. To perform simultaneous registrations of the surface oscillations and flow on the fluid surface the LED illumination under the bottom of the bath was synchronised with the even frames of video recording, while the LED illumination along the bath perimeter was synchronised with the odd frames. The experiments were conducted in a dark room in order to avoid a parasitic illumination.

The timeline of the experiment was as follows. At the initial moment, the fluid was at rest and we began to excite surface waves by applying sinusoidal signals of the same amplitude to the both subwoofers. The phase shift between the signals for subwoofers was equal to $\pi/2$. We note here, that due to small unintentional non-symmetry of the experimental setup and the vicinity of the exciting frequency to the resonance frequency, wave amplitudes $H_{1,2}(t)$ can be not exactly equal to each other and the phase difference $\psi(t)$ between waves can differ from $\pi/2$. At the moment of time $t^*= 1257 \, s$ the pumping was turned off and the fluid motion began to decay. The motion was recorded until a time $t_{end} = 1780 \, s$ when the wave and the vortex motion were damped to such an extent that they were no longer detectable. The analyses of the surface elevation $h(t,x,y)$ and of the two-dimensional velocity field on the fluid surface $(V_x(t,x,y), V_y(t,x,y))$ in the Fourier space with respect to spatial and temporal variables allows us to obtain the time-dependence of the wave amplitudes $H_1(t)$ and $H_2(t)$, their horizontal and vertical velocities on the surface of fluid, the phase difference $\psi(t)$ between excited waves, and the intensity of slow eddy currents $\Omega(x,y,0,t)$ on the fluid surface.

\begin{figure}[t]
%  \begin{picture}(300,270)
%    \put(-59,9.3){\includegraphics[width=0.53\linewidth]{picp1a.pdf}}
%        \put(-70,80){\rotatebox{90}{$\scriptstyle \text{vorticity}\ ||\Omega||, \ \ \text{1/s}$}}
%        %\put(198,80){\rotatebox{90}{$\scriptstyle \text{phase difference}\ \psi, \ \ \text{rad}$}}
%        \put(50,0){$\scriptstyle \text{time}\ t, \  \text{s}$}
%        \put(140,182){$\scriptstyle \varepsilon=0.33$}
%        \put(147,67){$\scriptstyle \varepsilon=0$}
%        \put(-75,200){a)}
%
%%        \put(-47.9,25){$|$}
%%        \put(-21.45,25){$|$}
%%        \put(-44.65,25){$|$}
%%        \put(-41.5,25){$|$}
%
%    \put(220,-15){\includegraphics[width=0.36\linewidth]{pic5.pdf}}
%        \put(290,0){$\scriptstyle \text{time}\ t, \  \text{s}$}
%        \put(295,188){$\scriptstyle \text{experiment}$}
%        \put(295,176){$\scriptstyle Ae^{-(t-t^\ast)/\tau_{exp}} + B$}
%        \put(220,80){\rotatebox{90}{$\scriptstyle \text{wave amplitude, arb.}$}}
%        \put(220,200){b)}
%  \end{picture}
  \includegraphics[width=\linewidth]{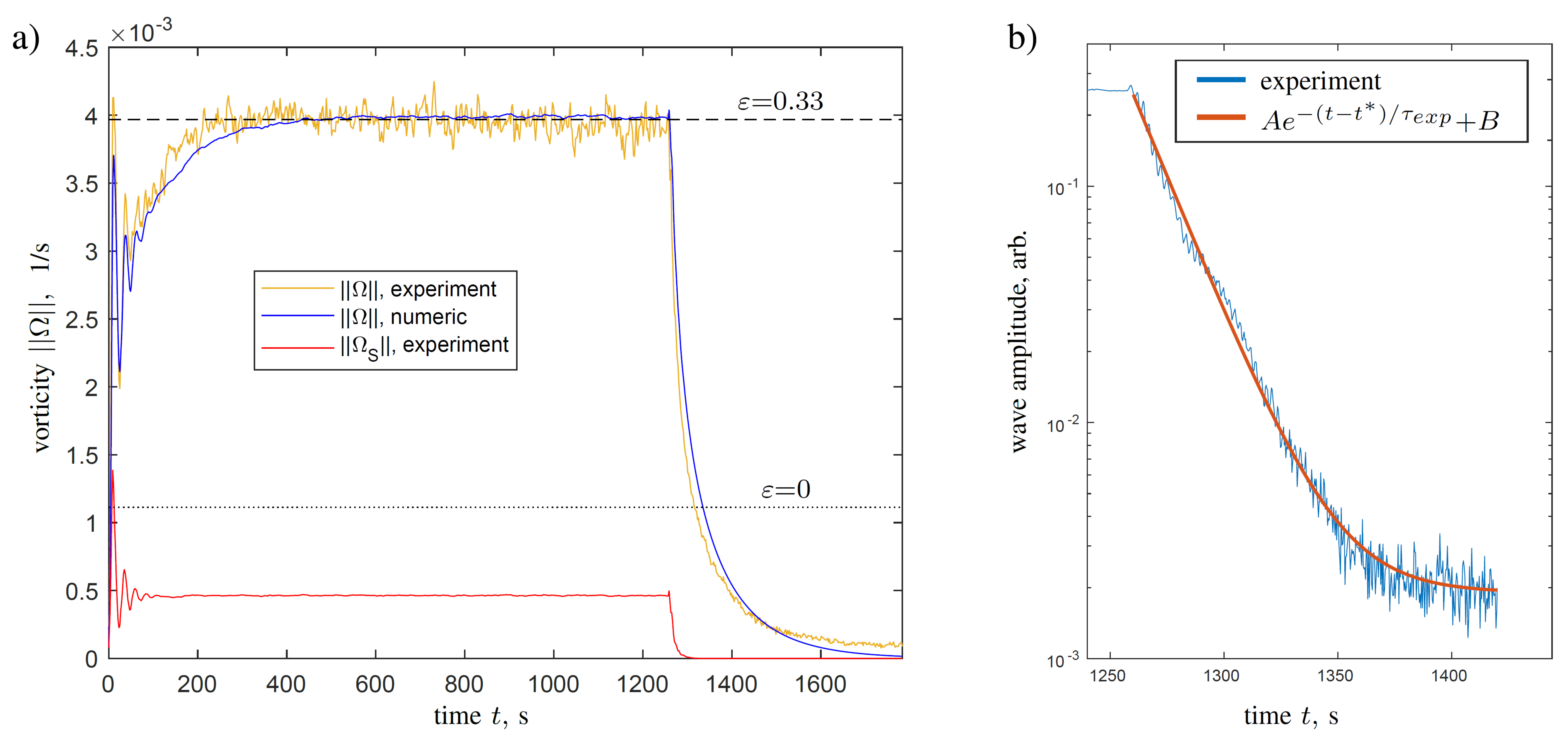}
  \caption{(a) The amplitude of eddy currents $||\Omega|| = ||\Omega_E|| + ||\Omega_S||$ depending on time for $38 \%$ glycerine-water solution. The yellow curve corresponds to the experimental data and the blue curve shows numerical result based on equation (\ref{eq:X}). The red curve demonstrates the dependence of Stokes contribution $||\Omega_S||$ on time based on the experimentally measured wave amplitudes $H_{1,2}(t)$ and $\psi(t)$. The horizontal lines show the time-asymptotic solutions, see expression (\ref{eq:W}), for the free surface case $\varepsilon=0$ and for the found parameter $\varepsilon=0.33$. (b) The decay of the wave amplitude after the pump is turned off.}
\label{pic:1}
\end{figure}

\begin{figure}[t]
%\noindent
%\begin{picture}(300,170)
%    \put(-107,-187){\includegraphics[width=0.6\linewidth]{pic2a.pdf}}% \\ (a)}
%    \put(20,0){$\scriptstyle \sqrt{t/t_E}$}
%    \put(-74,80){\rotatebox{90}{$\scriptstyle \text{vorticity} \ ||\Omega||, \ \text{1/s}$}}
%    \put(140,60){\rotatebox{90}{$\scriptstyle \text{phase difference \ }\psi, \text{\ \ rad}$}}
%    \put(-78,180){a)}
%
%    \put(165,1){\includegraphics[width=0.49\linewidth]{picp2b.pdf}}% \\ (b)}
%    \put(170,80){\rotatebox{90}{$\scriptstyle \text{vorticity} \ ||\Omega||, \ \text{1/s}$}}
%    \put(255,0){$\scriptstyle \sqrt{(t-t^\ast)/t_E}$}
%
%    \put(330,167){$\scriptstyle \varepsilon=0.33$}
%    \put(335,65){$\scriptstyle \varepsilon=0$}
%
%    \put(168,180){b)}
%\end{picture}
\includegraphics[width=\linewidth]{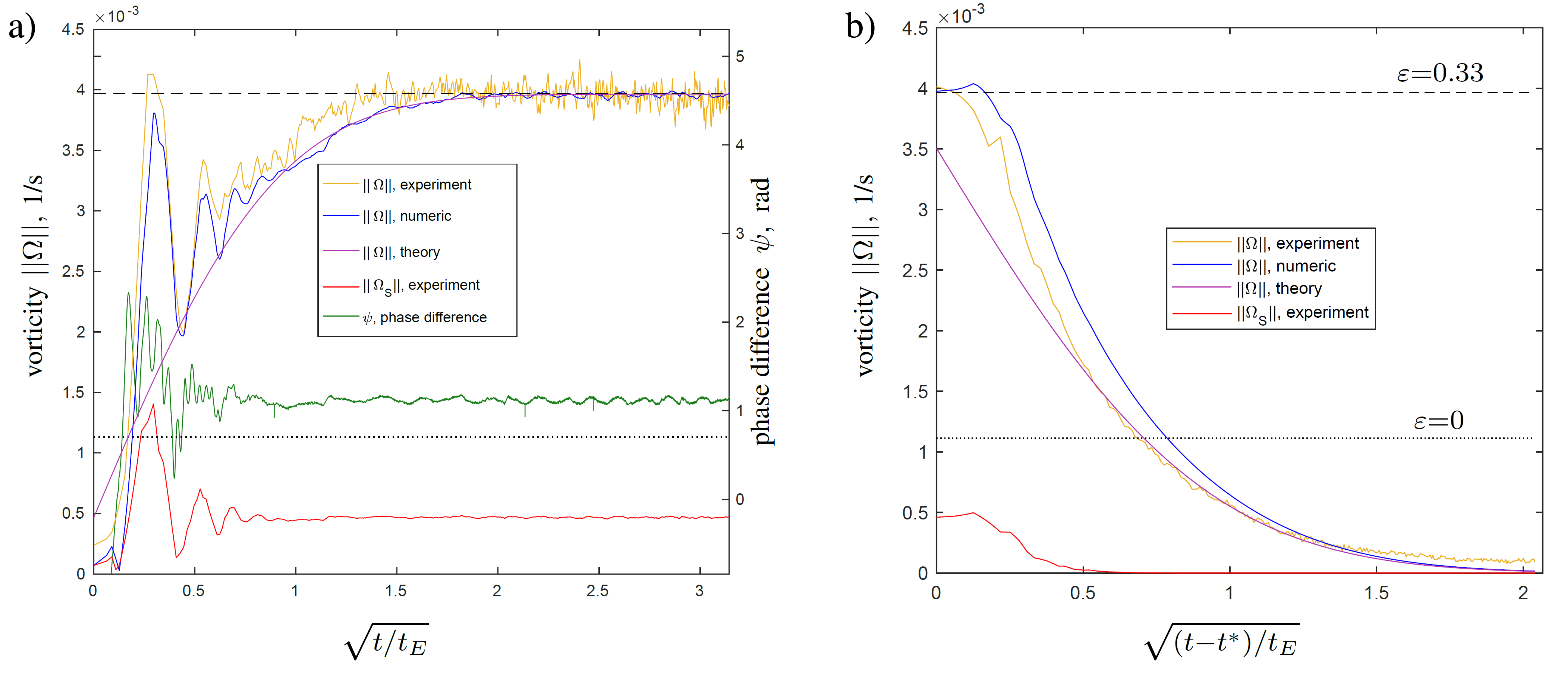}
\caption{Formation (a) and decay (b) of the eddy currents ($t_E \approx 125 \, s$) on the fluid surface for the same parameters as in Fig.~\ref{pic:1}. The yellow curves correspond to the experimental data, the blue curves show numerical results based on equation (\ref{eq:X}). The red curves demonstrate the Stokes drift contribution based on the experimentally measured wave amplitudes $H_{1,2}(t)$ and the phase difference $\psi(t)$. The violet curves show analytical results based on expressions (\ref{eq:13}) and (\ref{eq:16}).}
\label{pic:2}
\end{figure}

First, let us present the results which correspond to the low values of the effective Reynolds number (high fraction of glycerin and/or small pumping amplitude). In particular, Figs.~\ref{pic:1} and \ref{pic:2} demonstrate the behavior of $38 \%$ glycerine-water solution with the wave steepness $k H_1 = 5.4\times 10^{-3}$, $k H_2 = 4.8\times 10^{-3}$ in the regime of established wave motion. The solution has the fluid mass density $\rho = 1.091 \, g/cm^3$ and the kinematic viscosity $\nu = (3.1 \pm 0.05) \times 10^{-2} \, cm^2/s$ (measured by a viscosimeter). Fig.~\ref{pic:1}a shows the time-dependence of eddy currents $||\Omega||(t)$ together with the theoretical prediction. As before, the notion $||\dots||$ means that we drop the dependence on horizontal coordinates. For the slow eddy motion it can be easily restored: $\Omega(x,y,0,t) = -||\Omega|| \sin(kx) \sin(ky)$. As was explained in the previous section, the intensity of eddy currents can be described as a sum of Stokes drift and Euler contribution, i.e. $\Omega = \Omega_S + \Omega_E$. Both terms have the same dependence over horizontal coordinates (including the sign) and therefore $||\Omega|| = ||\Omega_S|| + ||\Omega_E||$. The dependence of the Stokes contribution on time is trivial, $||\Omega_S|| = e^{2kz} ||\Lambda(t)||$, and can be easily found by using expression (\ref{eq:6}), since we have measured $H_{1,2}(t)$ and $\psi(t)$, see the red curve in Fig.~\ref{pic:1}a. To obtain the Euler contribution one needs to solve numerically equation (\ref{eq:X}) for $||\Omega_E||$ with the same $||\Lambda(t)||$ in the right-hand side. Then the only unknown parameter is the compression modulus of the film, and varying the solution over this parameter and finding the best fit to the experimental data we obtain $\varepsilon = 0.33\pm0.02$, which corresponds to the blue curve in Fig.~\ref{pic:1}a. The horizontal lines on the same figure show the time-asymptotic solutions, see expression (\ref{eq:W}), for the free surface case $\varepsilon=0$ and for the found parameter $\varepsilon=0.33$. As it can be concluded, if the fluid surface was not contaminated then the excited eddy currents would be several times weaker provided that the wave motion would be the same. The effective Reynolds number for the considered case can be estimated as $\Omega/(\nu k^2) \approx 1$.

Next, we consider the formation and decay of the vortex motion for the same experimental conditions in more detail and compare the results with analytical predictions (\ref{eq:13}) and (\ref{eq:16}), which are reasonable because $t_E \approx 125 \, s$ is significantly longer than $t_S \approx 10 \, s$, see Fig.~\ref{pic:2}. The oscillations with period $\approx 24\, s$ visible on the formation curves, see Fig.~\ref{pic:2}a, correspond to the beating of the wave motion with the period being equal to the inverse difference between the frequencies of surface eigenmodes and of the pumping. The oscillations are not captured by our analytical result (\ref{eq:13}). With the exception of this fact, one can conclude that the agreement between the numerical, experimental, and analytical curves is quite good.
The phase difference $\psi(t)$ between standing waves in $X-$ and $Y-$ directions is shown in Fig.\ref{pic:2}a, see the green curve and the vertical axis on the right. The fast oscillations with period $\approx 5.3\, s$ at the beginning of the experiment are the result of excitation of neighboring modes characterized by $n=kL/\pi=7, 9$ when the pump is suddenly turned on. These modes do not make any contributions into $||\Lambda(t)||$, but disturb measured phase difference. Note also that the stationary value of $\psi$ is $\approx 1.1$ that differs from the phase shift of $\pi/2$ between electric signals applied to plunger drives due to inevitable non-symmetry of the experimental setup.

%The wave amplitudes $H_{1,2}(t)$ and the phase difference $\psi(t)$ used in the calculation of $||\Lambda(t)||$ were measured with a time step of $2\,s$ and with the same duration of averaging. Also, we have separately measured phase difference with time averaging over $1\,s$ and time step $1/12\,s$, see the green curve in Fig.\ref{pic:2}a. The fast oscillations with period $\approx 5.3\, s$ are the results of excitation of neighboring modes characterized by $n=kL/\pi=7, 9$ when the pump is suddenly turned on. These modes do not make any contributions into $||\Lambda(t)||$, but disturb measured phase difference $\psi(t)$ if the averaging time is relatively small. Note that the stationary value of $\psi$ is $\approx 1.1$ that differs from $\pi/2$. This means that the frequency difference between the operating $X$- and $Y$-eigenmodes is of the order of the resonance width $1/\tau$.

The decay process begins with a reduction of the pump oscillations to zero within one second. This stage and the subsequent evolution of the vorticity are presented in Fig.~\ref{pic:2}b. The agreement between experimental, numerical and analytical curves is also reasonable. Since the neighboring eigenmodes are again excited during the pump shutdown, measuring the phase difference $\psi(t)$ between the standing waves becomes a poorly defined task. To simplify, we assume that the phase shift $\psi(t)$ is constant and equals to its value in the stationary regime. This is justified, since the characteristic time of the phase change is of the order of the wave decay time. It is important to note that the Stokes drift contribution decays much faster than the Euler contribution and thus in our experiment we are able to see the eddy currents when the wave motion has already disappeared. This observation proves the existence of Euler contribution for the currents and demonstrates that the relation $t_E \gg t_S$ is valid.

The obtained value of the compression modulus of the film can be used to calculate the ratio of amplitudes of horizontal and vertical velocities on the fluid surface for the wave motion. By substituting $\varepsilon = 0.33$ in expression (\ref{eq:4}), one finds $||u_{\alpha}||/||u_z|| = 1.25$. The same ratio can be calculated using experimental data and we obtain $||u_x||/||u_z|| = 1.29\pm0.03$ for the wave propagating in the $X-$direction and $||u_y||/||u_z|| = 1.23\pm0.03$ for the wave propagating in the $Y-$direction. These values were obtained by averaging over time in the stationary regime, $400 \, s \leq t \leq 1200 \, s$. One can also study the damping of the wave motion based on the experimental dependencies $H_1(t)$ and $H_2(t)$ after the pumping is turned off. Fitting the dependencies by the law $A e^{-(t-t^*)/\tau_{exp}} + B$, where the constant $B$ corresponds to the level of noise in the measurement process ($B/A \lesssim 0.01$), we find $\tau_{exp} = (19.1 \pm 0.6) \, s$ for the $X-$wave and $\tau_{exp} = (21.5 \pm 0.8) \, s$ for the $Y-$wave, see Fig.~\ref{pic:1}b for illustration.  We believe that the difference between these two values is mainly due to the small asymmetry of the experimental setup. One can also obtain the decay time $\tau_{th}\approx 25\,s$ theoretically, using expression (\ref{eq:W''}) for the involved parameters. These results demonstrate that the theoretical model is reasonable despite its simplicity.

\begin{figure}[t]
%  \begin{picture}(300,200)
%    \put(0,0){\includegraphics[width=0.65\linewidth]{pic3.pdf}}
%    \put(5,90){\rotatebox{90}{$\scriptstyle \text{vorticity\ } ||\Omega||, \ \text{\ 1/s}$}}
%    \put(300,75){\rotatebox{90}{$\scriptstyle \text{large scale velocity\ }V_L, \text{\ cm/s}$}}
%    \put(150,0){$\scriptstyle \sqrt{t/t_E}$}
%
%    \put(155,50){\textcolor{dukeblue}{$\scriptstyle kH \approx 4.7 \times 10^{-3}$}}
%    \put(165,100){\textcolor{dukeblue}{$\scriptstyle kH \approx 1.1 \times 10^{-2}$}}
%  \end{picture}
  \includegraphics[width=0.8\linewidth]{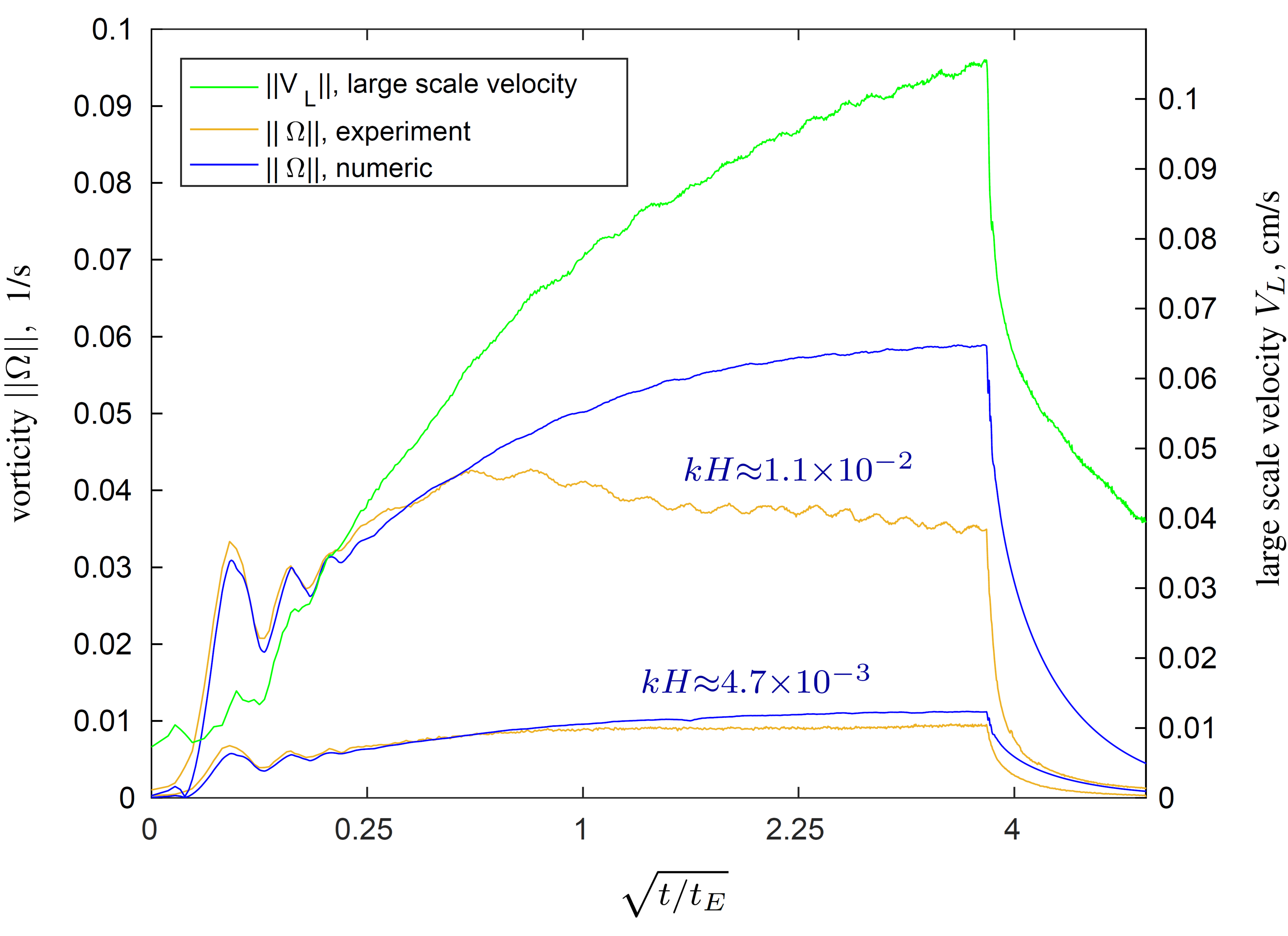}
  \caption{The time-dependence ($t_E \approx 320 \, s$) of the amplitude of eddy currents for $7.7 \%$ glycerine-water solution for two levels of excitation with the waves steepness $k H_{1,2}\approx 4.7 \times 10^{-3}$ and $kH_{1,2} \approx 1.1 \times 10^{-2}$. The compression modulus of the film $\varepsilon = 0.53\pm0.02$. The yellow curves correspond to the experimental data and the blue curves show numerical results based on equation (\ref{eq:X}). The green curve shows the rms of large-scale velocity $V_L$ of slow motion with the wave numbers $k_L \leq 0.26 \, cm^{-1}$ (vertical axis on the right) for the higher level of excitation.}
\label{pic:3}
\end{figure}

Next, let us turn to the results that correspond to the high values of the effective Reynolds number (low fraction of glycerin and/or large pumping amplitude). Fig.~\ref{pic:3} shows the behavior of $7.7 \%$ glycerine-water solution at different level of excitation with the waves steepness $k H_{1,2}\approx 4.7 \times 10^{-3}$ and $kH_{1,2} \approx 1.1 \times 10^{-2}$. The solution has the fluid mass density $\rho = 1.017 \, g/cm^3$ and the kinematic viscosity $\nu = (1.2 \pm 0.05) \times 10^{-2} \, cm^2/s$ (measured by a viscosimeter). The effective Reynolds numbers $\Omega/(\nu k^2)$ are approximately equal to $6$ and $27$ correspondingly. As one can see, the dynamic of the eddy currents substantially deviates from the theoretical prediction at times $t > 200 \, s$. This happens because the advection time $t_{\mathrm{adv}}$ becomes comparable with the viscous diffusion time $t_E \approx 320 \, s$. However, the advection is determined primarily not by the vortex lattice itself, but by the parasitic large-scale contribution ${\bm V}_L$ into the slow current, which was recently observed experimentally at large Reynolds numbers and at large times \cite{filatov2017formation}.

To be more precise, let us remind that equations (\ref{eq:X0}) and (\ref{eq:X1}) were obtained for the small values of the effective Reynolds number, $\mathrm{Re} = \Omega/(\nu k^2) \lesssim 1$, and they take into account only the second-order hydrodynamic nonlinearity (pair interaction between excited surface waves). When the condition is violated, one should add additional higher-order terms into these equations. In particular, we should add the contribution $(\bm V_E \nabla) \Omega_E$ in the left-hand side of expression (\ref{eq:X0}), which describes the advection of the vertical vorticity $\Omega_E$ by the velocity field ${\bm V}_E$. The velocity field ${\bm V}_E$ can be decomposed into the contribution of the vortex lattice ${\bm V}_\Omega$ and the parasitic large-scale velocity ${\bm V}_L$. Note that ${\bm V}_\Omega$ produces the main contribution into $\Omega_E$ due to $V_L\lesssim V_\Omega$ in our experiments and $kL\gg1$. Nevertheless, the first contribution $(\bm V_\Omega \nabla) \Omega_E$ is suppressed, since the vorticity $\Omega_E$ almost does not change in the direction of the velocity field ${\bm V}_\Omega$, and it turns out that the advection by the large-scale contribution, $(\bm V_L \nabla) \Omega_E$ plays the key role and defines the advection time.

%The term \textcolor{dukeblue}{$({\bm V}_L\nabla)\Omega$} in curl of Navier-Stokes equation produces more effective advection in comparison with the contribution \textcolor{dukeblue}{$({\bm V}_\Omega \nabla)\Omega$}, which is actually suppressed due to geometry of the velocity field \textcolor{dukeblue}{${\bm V}_\Omega$}, corresponding to the vortex lattice, since the vorticity  $\Omega$ almost does not change in the direction of the velocity field \textcolor{dukeblue}{${\bm V}_\Omega$}.

The green curve in Fig.~\ref{pic:3} shows the time-dependence of root mean square velocity $V_L$ of large-scale slow motion obtained by the Fourier filtering the velocity field of eddy currents for the wave numbers $k_L \leq 0.26 \, cm^{-1}$ (vertical axis on the right). The suppression of the amplitude of eddy currents started when the advection time by the large-scale velocity $t_{\mathrm{adv}} \sim 2 \pi/(k V_L)$ becomes smaller than the settling time of the vortex lattice $t_E$. For the involved parameters, one obtains $V_L \gtrsim 0.05 \, cm/s$ that qualitatively corresponds to the presented results. Note that the large-scale velocity $V_L$ decays more slowly after the pumping is turned off as compared to the dynamics of the vortex lattice. The blue curves in Fig.~\ref{pic:3} demonstrate the numerical solution of equation (\ref{eq:X}), where the compression modulus of the film was determined by finding the best fit in the range $0 \leq t \leq 80 \ s$, when the advection is negligible.

\begin{figure}[t]
%    \begin{picture}(300,190)
%        \put(-60,13){\includegraphics[width=0.425\linewidth]{pic4a.pdf}}% \\ (a)}
%        \put(-60,105){\rotatebox{90}{$\scriptstyle |\!|u_\alpha |\!|/|\!|u_z|\!|$}}
%        \put(20,10){$\scriptstyle 1/\sqrt{\varepsilon^2-\varepsilon\sqrt{2}+1}$}
%        \put(-65,205){a)}
%
%        \put(170,12.5){\includegraphics[width=0.41\linewidth]{pic4b.pdf}}% \\ (b)}
%        \put(255,10){$\scriptstyle \tau_{th}, \ \text{seconds}$}
%        \put(165,105){\rotatebox{90}{$\scriptstyle \tau_{exp}, \ \text{seconds}$}}
%        \put(160,205){b)}
%    \end{picture}
\includegraphics[width=\linewidth]{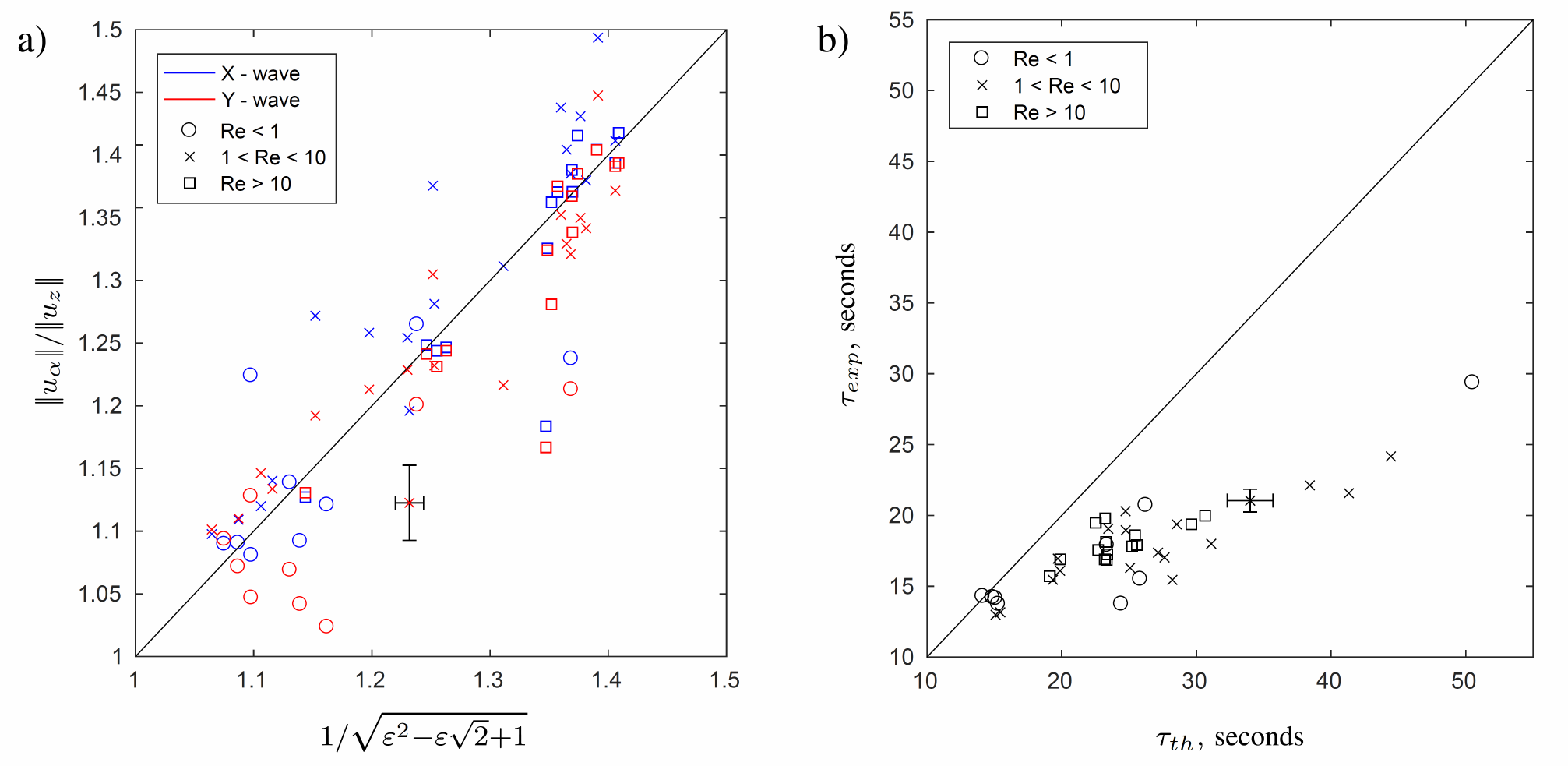}
\caption{(a) The comparison of the ratio of horizontal and vertical velocities on the fluid surface for the wave motion measured directly (vertical axis) and calculated theoretically (horizontal axis), see expression (\ref{eq:4}). Different colors correspond to the waves in $X-$ and $Y-$directions. (b) The comparison of the wave decay time measured directly after the pumping is switched off (vertical axis) and calculated theoretically (horizontal axis) according to expression (\ref{eq:W''}). The experimental values correspond to the largest decay time measured for $X-$ and $Y-$waves. Solid lines show the perfect agreement between theory and experiment. Error bars are approximately the same for all data points and thus are shown only for one of them to make the figure more readable.}
\label{pic:4}
\end{figure}

Finally, let us summarize the results obtained for different pumping amplitudes and for different glycerin concentrations. For each set of parameters, we calculated the effective Reynolds number $\Re = \Omega/(\nu k^2)$, where $\Omega$ is the maximum value of the measured vorticity, and the compression modulus $\varepsilon$ of the film by finding the best fit to the experimental results (as was discussed above). Then, by using expression (\ref{eq:4}), we found the ratio of amplitudes of horizontal and vertical velocities on the fluid surface for the wave motion. In Fig.~\ref{pic:4}a we compare the obtained values ($x-$axis) with the experimental results ($y-$axis). Different colors correspond to the waves in $X-$ and $Y-$directions. As one can see, in general the theory gives the correct trend, but the dispersion is quite large. Fig.~\ref{pic:4}b shows the comparison of the wave decay times measured directly after the pumping is switched off ($y-$axis) and calculated theoretically ($x-$axis), see expression (\ref{eq:W''}). The values on the $y-$axis correspond to the largest time of the decay times measured for the waves propagating in $X-$ and $Y-$directions. Despite this fact, our theory systematically underestimates the decay rate. We believe that this is caused by some other energy dissipation mechanisms, which were not taken into account. One of them is the friction of the fluid near the plunger which remained to be partially submerged into the fluid after the pumping was turned off.

\section{Conclusion}

It was established experimentally that eddy currents excited by crossed surface waves are the sum of the Stokes drift and Euler motion. We traced the separation into these two contributions during the dynamics of glycerine-water solution in the square cell. We started with the fluid at rest, then turned on the plungers exciting the surface waves, obtained steady flow, turned off the plungers and overwatched the decay process. The Euler contribution is excited by the waves but it is characterized by its own dynamics, whereas the Stokes drift is determined by instantaneous amplitudes of the waves. %\textcolor{dukeblue}{Note, that considered deep-water and low viscous limits lead to establishing of both finite Eulerian and Stokes contribution into slow currents, whereas ideal fluid with finite depth shows only Stokes drift and finite displacement of the surface \cite{mcallister2018set}. Here, we restricted ourselves to monochromatic orthogonal waves of extremely small amplitudes to keep small Reynolds numbers of the stationary eddy lattice. In nature, one has deal with stochastic wave field of larger amplitudes. In the case, the laminar flow grows up to turbulent one \cite{savelyev2012turbulence}, and the mechanism investigated in this work should be considered as one of the seed for the turbulent flow development~\cite{babanin2012numerical}.}

The difference between these two contributions was enhanced due to spontaneous formation of the surface film owing to fluid contamination. The wave frequency was $3\,Hz$ and the contamination typically has the strongest impact on the wave motion at these frequencies \cite{campagne2018impact}. In Ref.~\cite{parfenyev2018influence}, we modeled the contamination by assuming that the fluid surface is covered by a thin insoluble liquid (with zero shear elasticity) film. First, the presence of a film leads to the fact that the time scale of the dynamics of surface waves becomes much smaller than the time scale of the dynamics of Euler contribution to eddy currents. Second, it increases the amplitude of the Euler contribution, whereas the amplitude of the Stokes contribution remains unchanged as compared to the free surface case.

Now we found experimentally that the intensity of the eddy currents on the fluid surface exceeds theoretical predictions obtained for the free surface case, see Fig.~\ref{pic:1}. The result is in agreement with earlier measurements \cite{filatov2016nonlinear, filatov2016generation}, see also the discussion of these papers in Ref.~\cite[Sec.~VI.B]{parfenyev2018influence}. We extended the theory presented in Ref.~\cite{parfenyev2018influence} to the non-stationary case and applied it to describe the results of transient measurements, which include the stages of formation, steady-state, and decay of eddy currents. Our theory has the only free parameter --- the dimensionless compression modulus $\varepsilon$ of the surface film --- and it describes experimental results quantitatively, see Fig.~\ref{pic:2}. Moreover, the surface film is known to modify the wave motion. For example, it changes the ratio of horizontal and vertical velocities on the fluid surface and increases the wave damping, see Ref.~\cite[Sec.~VI.A]{parfenyev2018influence}. We have calculated both these quantities (\ref{eq:4}) and (\ref{eq:W''}) and found that the obtained value of the film compression modulus $\varepsilon$ leads to the values which are in a reasonable agreement with the experimental data, see Fig.~\ref{pic:4}. Despite the fact that all measurements in our experiment were carried out on the fluid surface, the theory also describes the motion in the bulk, and it shows that the film on the surface is capable of significantly influencing the fluid flow in the volume.

%The presented theory has a significant drawback --- it is non-universal and can be applied only to a special class of surface films.

The presented theory is applicable to a special class of compressible surface films, which can be characterized by the only rheological parameter --- the compression modulus. Recently, it was shown experimentally that viscoelastic films with different properties (non-zero shear elasticity) lead to the opposite effect --- they suppress the intensity of eddy currents \cite{francois2015inhibition}. Therefore, surface films have a potential to control the intensity of currents induced by the wave motion in a wide range. We hope that our results will motivate further work in this direction.

%\textcolor{red}{The phenomenon requires additional theoretical and experimental investigation. On the theoretical side, it is necessary to expand the model taking into account other rheological properties of the films. From the experimental point of view, one should study the penetration of the currents into the fluid bulk and collect the data for different films. Ways to solve these problems are known \cite{langevin2014rheology, xia2011upscale} and we hope that our results will motivate further work in this direction.}

\acknowledgments

Theoretical part of this work was supported by the Foundation for the Advancement of Theoretical Physics and Mathematics ''BASIS''. The experiment was conducted with the support of the Russian Science Foundation, Grant No. 17-12-01525.

\appendix
\section{Diffusion in a half-space}\label{sec:AppA}

Let us consider the diffusion in a half-space $z \leq 0$ with a fixed flux through the boundary $z=0$. The evolution of concentration $n (\bm r, t)$ is governed by the equations:
\begin{equation}\label{eq:app1}
 \partial_t n - D \nabla^2 n = 0, \quad \partial_z n\vert_{z=0} = S, \quad n\vert_{z\rightarrow -\infty} \rightarrow 0, \quad z \leq 0,
\end{equation}
where $D$ is a diffusion coefficient. The problem is equivalent to the following one:
\begin{equation}\label{eq:app2}
 \partial_t n - D \nabla^2 n = 2 D S \delta(z), \quad \partial_z n\vert_{z=0} = 0, \quad n\vert_{z\rightarrow -\infty} \rightarrow 0, \quad z \leq 0.
\end{equation}
Indeed, the diffusion equations are the same in the region $z<0$. We can integrate equation (\ref{eq:app2}) over the region $-\epsilon<z<0$ and by taking a limit $\epsilon \rightarrow 0$ one finds that $\partial_z n\vert_{z=-0} = S$. Note that we get the factor $1/2$ because we only integrate half of the Dirac delta function $\delta(z)$.

The diffusion in a half-space with a reflected boundary that corresponds to the condition $\partial_z n\vert_{z=0} = 0$ is equivalent to the diffusion in a whole space, but the initial condition $n(\bm r, 0) = n_0(\bm r)$ must be symmetrically reflected from the plane $z=0$. Therefore, we obtain the following problem
\begin{equation}\label{eq:app3}
  \partial_t n - D \nabla^2 n = 2 D S \delta(z), \quad n\vert_{z\rightarrow \pm \infty} \rightarrow 0, \quad -\infty<z<+\infty,
\end{equation}
which has to be supplemented by the initial condition $n(x,y,z,0) = n_0(x,y,z)+n_0(x,y,-z)$. This problem is equivalent to the original problem (\ref{eq:app1}) and can be solved instead of it.

%\bibliography{kin_vort}

%

\end{document}